\documentclass[thmsa]{article}
\usepackage{amsfonts}

\usepackage{sw20mitp}

\input tcilatex
\begin{document}

\title{Additional aspects of the generalized linear-fractional branching process}
\author{Nicolas Grosjean, Thierry Huillet \\
Laboratoire de Physique Th\'{e}orique et Mod\'{e}lisation,\\
CNRS-UMR 8089 et Universit\'{e} de Cergy-Pontoise,\\
2 Avenue Adolphe Chauvin, 95302, Cergy-Pontoise, FRANCE\\
E-mail: Nicolas.Grosjean@u-cergy.fr, Thierry.Huillet@u-cergy.fr}
\maketitle

\begin{abstract}
We derive some additional results on the Bienyam\'{e}-Galton-Watson
branching process with $\theta -$linear fractional branching mechanism, as
studied in \cite{Sag}. This includes: the explicit expression of the limit
laws in both the sub-critical cases and the super-critical cases with finite
mean, the long-run behavior of the population size in the critical case,
limit laws in the super-critical cases with infinite mean when the $\theta $%
-process is either regular or explosive, results regarding the time to
absorption, an expression of the probability law of the $\theta $-branching
mechanism involving Bell polynomials, the explicit computation of the
stochastic transition matrix of the $\theta -$process, together with its
powers.\newline

\end{abstract}

\section{Introduction}

Recently, in \cite{Sag}, a family of branching mechanisms involving
explosions was introduced: the so-called $\theta $-linear fractional family.
It fixes the reproduction law of some specific Bienyam\'{e}-Galton-Watson
branching processes \cite{Harris}, and it is given in terms of its
probability generating function (pgf). This pgf family has the remarkable
invariance under iterated composition property so that in principle the law
of the population size at each generation can be computed. This family
extends the classical linear-fractional model (obtained when $\theta =1$)
whose study dates back to Schr\"{o}der, (\cite{Harris}, p. $9$\ and \cite
{Schro}).\emph{\ }This makes computation of important statistical quantities
of great interest quite explicit. In this construction $\theta \in \left[
-1,1\right] $, with very special properties for the cases $\theta \in
\left\{ -1,0,1\right\} $ when $\theta $ is an integer. We shall revisit this 
$\theta $-family and give some additional results, among which:

- the expression of the limit laws in the subcritical cases and
super-critical cases with finite mean, solving respectively the associated
Schr\"{o}der and Poincar\'{e} functional equations.

- the long-run behavior of the population size in the critical case.

- limit laws in the super-critical cases with infinite mean when either the $%
\theta $-process is regular or explosive.

- information on the time to absorption defined as the infimum of the times
to extinction and explosion.

- an expression of the probability mass distribution of the $\theta $%
-branching mechanism, alternative to the one given in Proposition $4$ of 
\cite{Sag}, using of Faa di Bruno formulae and Bell polynomials.

- the explicit computation of the stochastic transition matrix of the
associated Bienyam\'{e}-Galton-Watson $\theta $-branching processes,
together with its powers. This gives some access to the resolvent of such
processes as a key ingredient to compute passage time statistics, hitting
probabilities,...

We end up this work by a short section of examples where the following
problem of concrete interest is addressed: what is the probability that,
given the $\theta $-branching process has not yet gone extinct at some given
generation, its extinction time be infinite with a large probability close
to $1$. We do some computations in the special cases $\theta \in \left\{
-1,0,1\right\} $.

\section{Generalities on Bienyam\'{e}-Galton-Watson (BGW) branching processes
}

We start with generalities on such BGW processes, including the case
displaying finite-time explosion, \cite{Sag}.

\subsection{The pgf approach}

Consider a discrete-time Bienyam\'{e}-Galton-Watson branching process \cite
{Harris} whose reproduction law is given by the (sub-)probability law $%
\mathbf{P}\left( M=m\right) =:\pi \left( m\right) $, $m\geq 0$ for the
number $M$ of offspring per capita. We assume $\pi \left( 0\right) >0$ so
that the process can go extinct. We let $\phi \left( z\right) =\mathbf{E}%
\left( z^{M}\right) =\sum_{m\geq 0}\pi \left( m\right) z^{m}$ be the
probability generating function of $M$ and we assume $\phi \left( 1\right)
\leq 1$.

With $N_{n}\left( 1\right) $\ the number of individuals alive at generation $%
n$\ given $N_{0}=1$, we have 
\begin{equation}
\mathbf{E}\left( z^{N_{n}\left( 1\right) }\right) :=\mathbf{E}\left(
z^{N_{n}}\mid N_{0}=1\right) =\phi ^{\circ n}\left( z\right) ,  \label{1.1}
\end{equation}
where $\phi ^{\circ n}\left( z\right) $\ is the $n$-th composition of $\phi
\left( z\right) $\ with itself, \footnote{%
Throughout this work, a pgf will therefore be a function $\phi $ which is
absolutely monotone on $\left( 0,1\right) $ with all nonnegative derivatives
of any order there, obeying $\phi \left( 1\right) \leq 1.$}. Similarly, if $%
N_{n}\left( i\right) $\ is the number of individuals alive at generation $n$%
\ given there are $N_{0}=i$\ independent founders, we clearly get 
\begin{equation}
\mathbf{E}\left( z^{N_{n}\left( i\right) }\right) :=\mathbf{E}\left(
z^{N_{n}}\mid N_{0}=i\right) =\phi ^{\circ n}\left( z\right) ^{i}.
\label{1.1.a}
\end{equation}
We shall also let 
\[
\tau _{i,j}=\inf \left( n\geq 1:N_{n}=j\mid N_{0}=i\right) , 
\]
the first hitting time of state $j\neq i$\ given $N_{0}=i\neq 0$.\newline

- If $\phi \left( 1\right) <1$, there is a positive probability $1-\phi
\left( 1\right) =:\pi \left( \infty \right) $ that $M=\infty $ (explosion is
made possible even at the first branching step): following \cite{Sag}, we
shall speak of an explosive or non-regular process.\newline

- If $\phi \left( 1\right) =1$ (regular case), depending on $\mu :=\mathbf{E}%
\left( M\right) \leq 1$ (i.e. the (sub-)critical case ) or $\mu >1$
(supercritical case): the process $N_{n}\left( 1\right) $ goes extinct with
probability $1$ or goes extinct with probability $\rho <1$ where $\rho $ is
the smallest fixed point solution in $\left[ 0,1\right] $ to $\phi \left(
\rho \right) =\rho $, respectively. In the latter case, the distribution of
the time to extinction $\tau _{1,0}$ is given by 
\[
\mathbf{P}\left( \tau _{1,0}\leq n\right) =\mathbf{P}\left( N_{n}\left(
1\right) =0\right) =\phi ^{\circ n}\left( 0\right) , 
\]
and the process explodes with probability $\overline{\rho }:=1-\rho $, but
not in finite time: only state $\left\{ 0\right\} $ is absorbing. Clearly
also, if there are $i$ independent founders instead of simply $1$, 
\[
\mathbf{P}\left( \tau _{i,0}\leq n\right) =\mathbf{P}\left( N_{n}\left(
i\right) =0\right) =\phi ^{\circ n}\left( 0\right) ^{i}. 
\]
\newline

- If $\phi \left( 1\right) <1$ (explosive case), $\mu :=\mathbf{E}\left(
M\right) =\infty $ because there is a positive probability $1-\phi \left(
1\right) $ that $M=\infty $. Notice that 
\[
\phi ^{\prime }\left( 1\right) =\mathbf{E}\left( M\cdot \mathbf{1}_{\left\{
M<\infty \right\} }\right) =\sum_{m\geq 1}m\pi \left( m\right) , 
\]
if this quantity exists (is finite). If $\phi \left( 1\right) <1$, state $%
\left\{ \infty \right\} $ should be added to the state-space $\Bbb{N}%
_{0}=\left\{ 0,1,...\right\} $ of $N_{n}\left( i\right) $ and then both
states are $\left\{ 0,\infty \right\} $ are absorbing. In this supercritical
case, $\rho <1$ always, and both the time to extinction $\tau _{1,0}$ and
the time to explosion $\tau _{1,\infty }$ of $N_{n}\left( 1\right) $ are
finite with positive probability, now with 
\begin{equation}
\left\{ 
\begin{array}{c}
\mathbf{P}\left( \tau _{1,0}\leq n\right) =\mathbf{P}\left( N_{n}\left(
1\right) =0\right) =\phi ^{\circ n}\left( 0\right) \stackunder{n\rightarrow
\infty }{\rightarrow }\rho =\mathbf{P}\left( \tau _{1,0}<\infty \right) . \\ 
\mathbf{P}\left( \tau _{1,\infty }>n\right) =\mathbf{P}\left( N_{n}\left(
1\right) <\infty \right) =\phi ^{\circ n}\left( 1\right) \stackunder{%
n\rightarrow \infty }{\rightarrow }\rho =\mathbf{P}\left( \tau _{1,\infty
}=\infty \right) .
\end{array}
\right.  \label{1.2}
\end{equation}
Thus $\rho $ and $\overline{\rho }$ are now also the probabilities that $%
\tau _{1,0}<\infty $ and $\tau _{1,\infty }<\infty $, respectively. We thus
have 
\begin{equation}
\left\{ 
\begin{array}{c}
\mathbf{P}\left( n<\tau _{1,0}<\infty \right) =\rho -\phi ^{\circ n}\left(
0\right) , \\ 
\mathbf{P}\left( n<\tau _{1,\infty }<\infty \right) =\overline{\rho }-\left(
1-\phi ^{\circ n}\left( 1\right) \right) =\phi ^{\circ n}\left( 1\right)
-\rho ,\text{ and,} \\ 
\mathbf{P}\left( n<\tau _{1}<\infty \right) =\mathbf{P}\left( 0<N_{n}\left(
1\right) <\infty \right) =\phi ^{\circ n}\left( 1\right) -\phi ^{\circ
n}\left( 0\right) ,
\end{array}
\right.  \label{1.3}
\end{equation}
where we defined the global absorption time $\tau _{1}:=\tau _{1,0}\wedge
\tau _{1,\infty }$. Clearly also, with $\tau _{i}:=\tau _{i,0}\wedge \tau
_{i,\infty }$%
\begin{equation}
\left\{ 
\begin{array}{c}
\mathbf{P}\left( n<\tau _{i,0}<\infty \right) =\rho ^{i}-\phi ^{\circ
n}\left( 0\right) ^{i}, \\ 
\mathbf{P}\left( n<\tau _{i,\infty }<\infty \right) =\left( 1-\rho
^{i}\right) -\left( 1-\phi ^{\circ n}\left( 1\right) ^{i}\right) =\phi
^{\circ n}\left( 1\right) ^{i}-\rho ^{i},\text{ and,} \\ 
\mathbf{P}\left( n<\tau _{i}<\infty \right) =\mathbf{P}\left( 0<N_{n}\left(
i\right) <\infty \right) =\phi ^{\circ n}\left( 1\right) ^{i}-\phi ^{\circ
n}\left( 0\right) ^{i}.
\end{array}
\right.  \label{1.3.a}
\end{equation}
\newline
Suppose a supercritical situation for which the extinction probability of $%
N_{n}\left( i\right) $ is smaller than $1$ (always the case if $\phi \left(
1\right) <1$). Of concrete interest is then the probability that, given the
process $N_{n}\left( i\right) $ has not yet gone extinct at generation $n$,
the extinction time of the process will be finite, namely 
\[
\mathbf{P}\left( \tau _{i,0}<\infty \mid N_{n}\left( i\right) >0\right) =%
\mathbf{P}\left( \tau _{i,0}<\infty \mid \tau _{i,0}>n\right) . 
\]
We get 
\[
\mathbf{P}\left( \tau _{i,0}<\infty \mid N_{n}\left( i\right) >0\right) =%
\frac{\rho ^{i}-\phi ^{\circ n}\left( 0\right) ^{i}}{1-\phi ^{\circ n}\left(
0\right) ^{i}}, 
\]
and the larger $n$, the smaller this probability because $\phi ^{\circ
n+1}\left( 0\right) >\phi ^{\circ n}\left( 0\right) $. There is thus a value 
$n_{c}$ of $n$ for which, with probability $c$ close to $1$, 
\begin{equation}
1-\mathbf{P}\left( \tau _{i,0}<\infty \mid N_{n_{c}}\left( i\right)
>0\right) =\frac{1-\rho ^{i}}{1-\phi ^{\circ n_{c}}\left( 0\right) ^{i}}=c%
\text{ (}=\text{say }0.99\text{).}  \label{neverext}
\end{equation}
This is the probability that some population with $i$\ founders, still alive
at generation $n_{c}$, will never go extinct.

\subsection{The transition matrix approach}

A Bienaym\'{e}-Galton-Watson process is a time-homogeneous Markov chain with
denumerable state-space $\Bbb{N}_{0}:=\left\{ 0,1,...\right\} $. Its
stochastic transition matrix is $P$, with entries $P\left( i,j\right)
=\left[ z^{j}\right] \phi \left( z\right) ^{i}=\mathbf{P}\left( N_{1}\left(
i\right) =j\right) $ (with $\left[ z^{j}\right] \phi \left( z\right) ^{i}$\
meaning the $z^{j}$-coefficient of the pgf $\phi \left( z\right) ^{i}$).
When there is explosion and in the supercritical cases, an interesting
problem arises when conditioning $N_{n}$ either on extinction or on
explosion. This may be understood as follows:\newline

The harmonic column vector $\mathbf{h}$, solution to $P\mathbf{h}=\mathbf{h}$%
, is given by its coordinates $h\left( i\right) =\rho ^{i}$, $i\geq 0$,
because $\sum_{j\geq 0}P\left( i,j\right) \rho ^{j}=\phi \left( \rho \right)
^{i}=\rho ^{i}$. Letting $D_{\mathbf{h}}:=$diag$\left( h\left( 0\right)
,h\left( 1\right) ,...\right) $, introduce the stochastic matrix $P_{\mathbf{%
h}}$ given by a Doob transform (\cite{Norris} and (\cite{RogWil}, p. $327$): 
$P_{\mathbf{h}}=D_{\mathbf{h}}^{-1}PD_{\mathbf{h}}$ or $P_{\mathbf{h}}\left(
i,j\right) =h\left( i\right) ^{-1}P\left( i,j\right) h\left( j\right)
=P\left( i,j\right) \rho ^{j-i}$, $i,j\geq 0$. Note $h\left( N_{n}\left(
i\right) \right) =\rho ^{N_{n}\left( i\right) }$ is a martingale because $%
\mathbf{E}\left( h\left( N_{n}\left( i\right) \right) \right) =\phi ^{\circ
n}\left( \rho \right) ^{i}=\rho ^{i}=h\left( i\right) =h\left( N_{0}\left(
i\right) \right) $. Then $P_{\mathbf{h}}$ is the transition matrix of $N_{n}$
conditioned on almost sure extinction. Equivalently, when conditioning $N_{n}
$ on almost sure extinction, one is led to a regular subcritical BGW process
with new branching mechanism $\phi _{0}\left( z\right) =\phi \left( \rho
z\right) /\rho $, satisfying $\phi _{0}\left( 1\right) =1$ and $\phi
_{0}^{\prime }\left( 1\right) =\phi ^{\prime }\left( \rho \right) <1$.
Indeed, $\phi _{0}\left( z\right) =\sum_{j\geq 0}P_{\mathbf{h}}\left(
1,j\right) z^{j}$. Upon iterating, we get $\phi _{0}^{\circ n}\left(
z\right) =\phi ^{\circ n}\left( \rho z\right) /\rho $.\newline

Similarly, when conditioning $N_{n}$ on almost sure explosion, one is led to
an explosive supercritical BGW process with new Harris-Sevastyanov branching
mechanism $\phi _{\infty }\left( z\right) =\left[ \phi \left( \rho +%
\overline{\rho }z\right) -\rho \right] /\overline{\rho }$, satisfying $\phi
_{\infty }\left( 0\right) =0$ and $\phi _{\infty }\left( 1\right) =\left(
\phi \left( 1\right) -\rho \right) /\overline{\rho }<1$. Upon iterating, we
have $\phi _{\infty }^{\circ n}\left( z\right) =\left[ \phi ^{\circ n}\left(
\rho +\overline{\rho }z\right) -\rho \right] /\overline{\rho }$.\newline

The second largest eigenvalue of $P$ is $\gamma =\phi ^{\prime }\left( \rho
\right) <1$. The corresponding eigenvector $\mathbf{u}$ obeys $P\mathbf{u}%
=\gamma \mathbf{u}$ with $u\left( i\right) =i\rho ^{i-1}$, $i\geq 1$,
because $\sum_{j\geq 1}P\left( i,j\right) j\rho ^{j-1}=\phi ^{\prime }\left(
\rho \right) i\phi \left( \rho \right) ^{i-1}=\gamma i\rho ^{i-1}$.
Conditioning $N_{n}$ on never hitting $\left\{ 0,\infty \right\} $ in the
remote future is given by the $Q$-process with stochastic transition matrix $%
Q=\gamma ^{-1}D_{\mathbf{u}}^{-1}PD_{\mathbf{u}}$ or $Q\left( i,j\right)
=\gamma ^{-1}u\left( i\right) ^{-1}P\left( i,j\right) u\left( j\right)
=\gamma ^{-1}\rho ^{j-i}i^{-1}P\left( i,j\right) j$, $i,j\geq 1$ (see \cite
{Lamb} and \cite{Sag}, Section $6$ in the $\theta $-special case).\newline

There are classes of discrete branching processes for which the pgf $\phi
^{\circ n}\left( z\right) $ of $N_{n}\left( 1\right) $ is exactly
computable, thereby making the above computations concrete and somehow
explicit.

\section{The $\theta $-linear fractional branching mechanism model, 
\protect\cite{Sag}}

With $\left| \theta \right| \leq 1$, $a,b>0$ and $z_{c}\geq 1$, we shall
consider the $\theta $-linear fractional branching mechanism model, namely, 
\begin{equation}
\left\{ 
\begin{array}{c}
\phi \left( z\right) =z_{c}-\left( a\left( z_{c}-z\right) ^{-\theta
}+b\right) ^{-1/\theta }\text{ or} \\ 
\left( z_{c}-\phi \left( z\right) \right) ^{-\theta }=a\left( z_{c}-z\right)
^{-\theta }+b,
\end{array}
\right.   \label{2.1}
\end{equation}
and for those values of $z_{c}\geq 1$ and $a,b>0$ for which $\phi $ is a pgf
with $\phi \left( 1\right) \leq 1$. The case $\theta =0$ will be considered
in (\ref{2.4}).

\subsection{The boundary cases $\theta =\pm 1$}

The boundary cases $\theta =\pm 1$ deserve a special treatment that we shall
first evacuate.\newline

$\bullet $ When $\theta =1$, $\phi \left( z\right) =z_{c}-\left( a\left(
z_{c}-z\right) ^{-1}+b\right) ^{-1}$ is an homographic map. Assuming $a+b>1$
and introducing the probabilities $p_{0}=1/\left( a+b\right) $, $q=a/\left(
a+b\right) $, with $p_{0}+q_{0}=1$ and $p+q=1$, this is also ($a=q/p_{0}$, $%
b=p/p_{0}$) 
\[
\frac{1}{z_{c}-\phi \left( z\right) }=\frac{q}{p_{0}}\frac{1}{z_{c}-z}+\frac{%
p}{p_{0}}. 
\]
Note $\phi \left( z_{c}\right) =z_{c}$ but $z_{c}$ is not the convergence
radius of $\phi $, which is $z_{c}+q/p$.\newline

- In the particular case $z_{c}=1$, we have the two following
interpretations for $\phi \left( z\right) :$

\begin{proposition}
$\left( i\right) $ When $z_{c}=1$, $\phi \left( z\right) =q_{0}+p_{0}\left(
qz\right) /\left( 1-pz\right) $, the classical form of the simple linear
fractional model. This pgf is the one of a random variable $M$\ obtained as $%
M\stackrel{d}{=}G\cdot B$ (equality in law), where $B$\ is Bernoulli$\left(
p_{0}\right) $\ distributed, independent of $G$, a geometric$\left(
1/q\right) $\ distributed random variable.

$\left( ii\right) $ When $z_{c}=1$ and if $b<1$, we also have 
\[
\phi \left( z\right) =\frac{1+\left( 1-z\right) \frac{b-1}{a}}{1+\left(
1-z\right) \frac{b}{a}},
\]
which can be put in the alternative form 
\[
\phi \left( z\right) =\frac{\beta \left( \beta _{0}+\alpha _{0}z\right) }{%
1-\alpha \left( \beta _{0}+\alpha _{0}z\right) },
\]
while defining the probabilities $\alpha _{0}=\left( 1-b\right) /a$, $\alpha
=b$\ and $\beta _{0}=1-\alpha _{0}$, $\beta =1-\alpha $. This $\phi \left(
z\right) $ is thus the pgf of the random variable 
\[
M\stackrel{d}{=}\sum_{k=1}^{G}B_{k},
\]
where $G$\ now is geometric$\left( 1/\beta \right) $\ distributed,
independent of the sequence of independent and identically distributed $%
\left( B_{k}\right) _{k\geq 1}$, with $B_{1}$\ Bernoulli$\left( \alpha
_{0}\right) $\ distributed. $M$\ is thus a Bernoulli-thinned version of $G$
in the sense of \cite{Steu}.\newline
\end{proposition}

We have $\mu :=\mathbf{E}\left( M\right) =\phi ^{\prime }\left( 1\right)
=p_{0}/q=1/a$ and 
\[
\left\{ 
\begin{array}{c}
\phi ^{\circ n}\left( z\right) =1-\left( a_{n}\left( 1-z\right)
^{-1}+b_{n}^{{}}\right) ^{-1}\text{ where} \\ 
a_{n}=a^{n}\text{ and }b_{n}=b\left( 1+a+...+a^{n-1}\right)
\end{array}
\right. . 
\]
Depending on $a>1$, $a=1$ or $a<1$, the corresponding branching process is
subcritical, critical or supercritical. In the supercritical case $%
a=q/p_{0}<1$ the extinction probability is $\rho =q_{0}/p<1$.\newline

- If now $z_{c}>1$, the additional constraints $\phi \left( 0\right) \in
\left( 0,1\right) $ and $\phi \left( 1\right) \leq 1$ impose $%
p_{0}<q+pz_{c}\leq p+p_{0}$. This family is of interest because its $n$-th
iterate is explicit, also homographic, with 
\begin{equation}
\left\{ 
\begin{array}{c}
\phi ^{\circ n}\left( z\right) =z_{c}-\left( a_{n}\left( z_{c}-z\right)
^{-1}+b_{n}^{{}}\right) ^{-1}\text{ where} \\ 
a_{n}=a^{n}\text{ and }b_{n}=b\left( 1+a+...+a^{n-1}\right) .
\end{array}
\right.  \label{2.1.1}
\end{equation}
Thus for instance, if $z_{c}>1$, and $q+pz_{c}<p+p_{0}$%
\[
\mathbf{P}\left( n<\tau _{1}<\infty \right) =\phi ^{\circ n}\left( 1\right)
-\phi ^{\circ n}\left( 0\right) =\frac{a_{n}}{\left( a_{n}+b_{n}z_{c}\right)
\left( a_{n}+b_{n}\left( z_{c}-1\right) \right) }, 
\]
with 
\[
\mathbf{P}\left( n<\tau _{1}<\infty \right) \stackunder{n\rightarrow \infty 
}{\sim }\left\{ 
\begin{array}{c}
\frac{\left( a-1\right) ^{2}}{\left( a-1+bz_{c}\right) \left( a-1+b\left(
z_{c}-1\right) \right) }a_{n}^{-1}\text{ if }a>1 \\ 
\frac{\left( a-1\right) ^{2}}{b^{2}z_{c}\left( z_{c}-1\right) }a^{n}\text{
if }a<1 \\ 
\frac{1}{b^{2}z_{c}\left( z_{c}-1\right) }n^{-2}\text{ if }a=1
\end{array}
\right. . 
\]
When $a=1$ ($p_{0}=$ $q$ and $q_{0}=$ $p$), the tails of $\tau _{1}$ are no
longer asymptotically geometric, rather they are power-law with tail index $%
2 $.\newline

$\bullet $ When $\theta =-1$, $\phi \left( z\right) =az+z_{c}\left(
1-a\right) -b$ is the affine map and, if $\phi \left( 1\right) =1$, the
corresponding branching process is the regular death process as each
individual can only either die or survive upon splitting. With $\pi \left(
1\right) =a$, $\pi \left( 0\right) =z_{c}\left( 1-a\right) -b=1-\pi \left(
1\right) $, $\phi \left( z\right) =\pi \left( 1\right) z+\pi \left( 0\right) 
$ and the corresponding branching process is subcritical, always, with mean $%
\mu =\pi \left( 1\right) =a<1$. With $\pi _{n}\left( 0\right) +\pi
_{n}\left( 1\right) =1$, we have 
\[
\phi ^{\circ n}\left( z\right) =\pi _{n}\left( 0\right) +\pi _{n}\left(
1\right) z\text{, where }\pi _{n}\left( 1\right) =\pi \left( 1\right) ^{n}. 
\]

If $\phi \left( 1\right) <1$, the corresponding branching process is an
explosive process where each individual can either die, survive or give
birth to infinitely many descendants on splitting. The additional
constraints $\phi \left( 0\right) \in \left( 0,1\right) $ and $\phi \left(
1\right) <1$ impose $\pi \left( 1\right) =a\in \left( 0,1\right) $, $\pi
\left( 0\right) =z_{c}\left( 1-a\right) -b<1-\pi \left( 1\right) =1-a$, thus 
$\left( z_{c}-1\right) \left( 1-a\right) <b$. This family is of interest
because its $n$-th iterate is again explicit 
\begin{equation}
\left\{ 
\begin{array}{c}
\phi ^{\circ n}\left( z\right) =z_{c}-\left( a_{n}\left( z_{c}-z\right)
+b_{n}\right) \text{ with} \\ 
a_{n}=a^{n}\text{ and }b_{n}=b\left( 1+a+...+a^{n-1}\right) =b\frac{1-a^{n}}{%
1-a}
\end{array}
\right.  \label{2.2}
\end{equation}
and again in the same class of affine maps. With $\pi _{n}\left( 0\right)
+\pi _{n}\left( 1\right) <1$, this is also 
\[
\phi ^{\circ n}\left( z\right) =\pi _{n}\left( 0\right) +\pi _{n}\left(
1\right) z\text{, where }\pi _{n}\left( 0\right) =\pi \left( 0\right) \frac{%
1-a^{n}}{1-a}\text{ and }\pi _{n}\left( 1\right) =\pi \left( 1\right) ^{n}. 
\]
We have 
\begin{eqnarray*}
\mathbf{P}\left( N_{n}\left( 1\right) <\infty \right) &=&\pi _{n}\left(
0\right) +\pi _{n}\left( 1\right) =\pi \left( 0\right) \frac{1-a^{n}}{1-a}%
+\pi \left( 1\right) ^{n} \\
\stackunder{n\rightarrow \infty }{\rightarrow }\mathbf{P}\left( N_{\infty
}\left( 1\right) <\infty \right) &:&=\pi \left( 0\right) /\left( 1-a\right)
<1.
\end{eqnarray*}
$\mathbf{P}\left( N_{n}\left( 1\right) =\infty \right) $ is an increasing
sequence. The relative rate of approach of $\mathbf{P}\left( N_{n}\left(
1\right) =\infty \right) $ to its limiting value decays geometrically with 
\[
\frac{\mathbf{P}\left( N_{\infty }\left( 1\right) =\infty \right) -\mathbf{P}%
\left( N_{n}\left( 1\right) =\infty \right) }{\mathbf{P}\left( N_{\infty
}\left( 1\right) =\infty \right) }=a^{n}. 
\]
Note $\mathbf{P}\left( n<\tau _{1}<\infty \right) =\phi ^{\circ n}\left(
1\right) -\phi ^{\circ n}\left( 0\right) =a^{n}$, an exact geometric
distribution.

\subsection{The case $\theta \in \left( -1,1\right) $}

Although we deal here with the case $\theta \in \left( -1,1\right) $, we,
somehow abusively, extend the range of the parameter set to its boundary
whenever it causes no particular problem.\newline

- With $\theta \in \left( -1,1\right) $, $a,b>0$ and $z_{c}=\sup \left(
z>0:\phi \left( z\right) <\infty \right) \geq 1$, let us reconsider $\phi
\left( z\right) $ as defined by (\ref{1.1}). Note now $\phi \left(
z_{c}\right) \leq z_{c}$ ($=z_{c}$ if $\theta \in \left( 0,1\right] $) and $%
z_{c}>1$ could produce $\phi \left( 1\right) <1$, the explosion opportunity.
This family is of interest because its $n$-th iterate is also explicit with
(if $\theta \neq 0$)

\begin{equation}
\left\{ 
\begin{array}{c}
\phi ^{\circ n}\left( z\right) =z_{c}-\left( a_{n}\left( z_{c}-z\right)
^{-\theta }+b_{n}\right) ^{-1/\theta }\text{ where} \\ 
a_{n}=a^{n}\text{ and }b_{n}=b\left( 1+a+...+a^{n-1}\right) ,
\end{array}
\right.  \label{2.3}
\end{equation}
and it is in the same class as $\phi $, although for a different set of
parameters $a,b$ (an invariance under iteration property).

The case $\theta =0$\ is defined by continuity from the case $\theta \in
\left( -1,1\right) \backslash \left\{ 0\right\} $\ while observing 
\begin{equation}
\phi \left( z\right) =z_{c}-\left( a\left( z_{c}-z\right) ^{-\theta }+\left(
1-a\right) \left( z_{c}-\rho \right) ^{-\theta }\right) ^{-1/\theta }%
\stackunder{\left| \theta \right| \rightarrow 0}{\rightarrow }z_{c}-\left(
z_{c}-\rho \right) ^{1-a}\left( z_{c}-z\right) ^{a},  \label{2.4}
\end{equation}
with $\phi \left( 1\right) <1$\ if $z_{c}>1$. Notice that if $z_{c}=1$, $%
\phi \left( 1\right) =1$\ and $\mu =\infty $\ (the only regular case with
infinite mean).\newline

There are three cases, depending on $\mu :=\mathbf{E}\left( M\right) <1$, $%
=1 $ or $>1$:\newline

$\bullet $ $\left( A\right) :$ subcritical cases:

$\left( i\right) $ If $\theta \in \left( 0,1\right] $, $z_{c}=1$, $a>1$, $%
b>0 $, then $\mu =a^{-1/\theta }<1$. Again, if $\theta =1$, $\phi \left(
z\right) =q_{0}+p_{0}qz/\left( 1-pz\right) $ with $p_{0}=1/\left( a+b\right) 
$, $p=b/\left( a+b\right) ,$ the classical form of the $1$-fractional model
as the composition of a Bernoulli$\left( p_{0}\right) $ pgf with the one of
a geometric$\left( p/q\right) $ pgf.

$\left( ii\right) $ If $\theta \in \left( -1,1\right] $, $z_{c}>1$, $a\in
\left( 0,1\right) $ and $b=\left( 1-a\right) \left( z_{c}-1\right) ^{-\theta
}$, then $\mu =a<1$.

$\left( iii\right) $ If $\theta =-1$, $z_{c}=1$ and $a\in \left( 0,1\right) $%
, then $\mu =a<1$.\newline

$\bullet $ $\left( B\right) :$ critical case ($\mu =1$): this situation
occurs only when $\theta \in \left( 0,1\right] $, $z_{c}=1$, $a=1$, $b>0$.%
\newline

$\bullet $ $\left( C\right) :$ supercritical case ($\infty \geq \mu >1$): $%
\theta \in \left( -1,1\right] $, $z_{c}\geq 1$, $a\in \left( 0,1\right) $, $%
b=\left( 1-a\right) \left( z_{c}-\rho \right) ^{-\theta }$ where
equivalently $\rho =z_{c}-\left( \left( 1-a\right) /b\right) ^{1/\theta }$
is the extinction probability of the process, as the smallest solution in
the interval $\left[ 0,1\right] $ to $\phi \left( \rho \right) =\rho $ with $%
\rho \in \left( 0,1\right) $. We have $a=\phi ^{\prime }\left( \rho \right) $%
.

In the supercritical case with $z_{c}>1$, then $\mu =\infty $ because in
this case, 
\[
\phi \left( 1\right) =z_{c}-\left( a\left( z_{c}-1\right) ^{-\theta }+\left(
1-a\right) \left( z_{c}-\rho \right) ^{-\theta }\right) ^{-1/\theta }<1 
\]
and $M=\infty $ with a positive probability.

In general, we have $\phi ^{\prime }\left( 1\right) =a\left( a+b\left(
z_{c}-1\right) ^{\theta }\right) ^{-\left( \theta +1\right) /\theta
}=a\left( a+\left( 1-a\right) \left( \frac{z_{c}-1}{z_{c}-\rho }\right)
^{\theta }\right) ^{-\left( \theta +1\right) /\theta }$ which coincides with 
$\mu $ if $z_{c}=1$. We conclude that in the supercritical case with $%
z_{c}=1 $%
\[
\mu =\left\{ 
\begin{array}{c}
\infty \text{ if }\theta \in \left( -1,0\right] \text{, }a\in \left(
0,1\right) \\ 
a^{-1/\theta }\text{ if }\theta \in \left( 0,1\right] \text{, }a\in \left(
0,1\right)
\end{array}
\right. . 
\]
In the first case,

- if $\theta \in \left( -1,0\right) $, $a\in \left( 0,1\right) $ then $\mu
=\infty $ as a result of finite-time explosion because $\phi \left( 1\right)
=1-\left( \left( 1-a\right) \left( 1-\rho \right) ^{-\theta }\right)
^{-1/\theta }<1$ (explosive case).

- if $\theta =0$, $a\in \left( 0,1\right) $, $\mu =\infty $ even though $%
\phi \left( 1\right) =1$ (the only regular case with infinite mean). \newline

\textbf{Remarks:}

$\left( i\right) $ To the subset of models $\left( A\right) $ to $\left(
B\right) $, we have added the special affine case $\theta =-1$ with $z_{c}=1$%
. If $z_{c}>1$, the affine model is supercritical with $\mu =\infty $
because the branching event $M=\infty $ has a positive probability. The
special case $\theta =0$ is supercritical with $\mu =\infty $ both when $%
z_{c}=1$ and $z_{c}>1$. The special case $\theta =1$ corresponds to the
standard linear fractional model and its criticality status has been
included in the above classification.

$\left( ii\right) $ Due to the invariance under iterated composition of the $%
\theta $-family of pgfs, it holds that $\left[ \phi ^{\circ n}\right]
^{-1}\left( z\right) =\phi ^{\circ \left( -n\right) }\left( z\right) $: the
inverse function of $\phi ^{\circ n}\left( z\right) $ simply is $\phi
^{\circ \left( -n\right) }\left( z\right) $, obtained while substituting $-n$
to $n$ in $\phi ^{\circ n}\left( z\right) $, (a time-reversal property).

\section{Limit laws}

We shall investigate different limit laws concerning cases $\left( A\right) $
to $\left( C\right) $.

\subsection{Limit laws (subcritical/critical and super-critical with finite
mean cases)}

$\bullet $ \textbf{Subcritical case with }$\mu <1$:

In the subcritical case, considering the population size, given it is
positive, gives rise to a limiting random variable as the generation number
goes to infinity. This limiting random variable is known as the
quasi-stationary Yaglom limit, \cite{Yag}.

In our context, there are three different cases where this situation can
occur:\newline

$\left( A\right) /\left( i\right) .$ In this case, with $\theta \in \left(
0,1\right] $, $z_{c}=1$, $a>1$, $b>0$ and $\phi \left( z\right) =1-\left(
a\left( 1-z\right) ^{-\theta }+b\right) ^{-1/\theta }$, $N_{n}\mid N_{n}>0%
\stackrel{d}{\rightarrow }N_{\infty }$ where $N_{\infty }$ is a random
variable with value in $\Bbb{N}_{0}=\left\{ 1,2,...\right\} $ whose pgf $%
\phi _{\infty }\left( z\right) :=\mathbf{E}\left( z^{N_{\infty }}\right)
=\sum_{l\geq 1}\pi _{\infty }\left( l\right) z^{l}$ obeys the Schr\"{o}der
functional equation 
\begin{equation}
\overline{\phi }_{\infty }\left( \phi \left( z\right) \right) =\mu \overline{%
\phi }_{\infty }\left( z\right) \text{, }\overline{\phi }_{\infty }\left(
z\right) =1-\phi _{\infty }\left( z\right) \text{, }\mu =a^{-1/\theta }.
\label{Schr}
\end{equation}
Note $\phi \left( z\right) =\overline{\phi }_{\infty }^{-1}\left( \mu 
\overline{\phi }_{\infty }\left( z\right) \right) $ and thus $\phi ^{\circ
n}\left( z\right) =\overline{\phi }_{\infty }^{-1}\left( \mu ^{n}\overline{%
\phi }_{\infty }\left( z\right) \right) $.

\begin{proposition}
With $\alpha =\frac{a-1}{a+b-1}$ and $\beta =\frac{b}{a+b-1}$ ($\alpha
+\beta =1$), we find the pgf of the Yaglom quasi-stationary limit $N_{\infty
}$ as 
\begin{equation}
\phi _{\infty }\left( z\right) =1-\frac{1-z}{\left( \alpha +\beta \left(
1-z\right) ^{\theta }\right) ^{1/\theta }},  \label{Yag}
\end{equation}
obeying $\phi _{\infty }\left( 0\right) =0$, $\phi _{\infty }\left( 1\right)
=1$ and with mean $\mu _{\infty }:=\phi _{\infty }^{\prime }\left( 1\right)
=\alpha ^{-1/\theta }=\left( \frac{a-1}{a+b-1}\right) ^{-1/\theta }$.
\end{proposition}

If in particular $\theta =1$, 
\[
\phi _{\infty }\left( z\right) =\frac{z}{1+\frac{\beta }{\alpha }\left(
1-z\right) }=\frac{\alpha z}{1-\beta z} 
\]
is the pgf of a geometric random variable with mean $1+\beta /\alpha
=1/\alpha $. Thus $\pi _{\infty }\left( l\right) =\mathbf{P}\left( N_{\infty
}=l\right) =\alpha \beta ^{l-1}$, $l\geq 1$, decays geometrically fast.

\begin{corollary}
Defining $\overline{\pi }_{\infty }\left( k\right) :=\sum_{l>k}\pi _{\infty
}\left( l\right) $, 
\[
\overline{\pi }_{\infty }\left( k\right) \stackunder{k\uparrow \infty }{\sim 
}-\frac{1}{\Gamma \left( -\theta \right) }\frac{\beta }{\theta \alpha
^{1+1/\theta }}k^{-\left( 1+\theta \right) },
\]
displaying power law tails with index $1+\theta $ if $\theta \in \left(
0,1\right) :$ $N_{\infty }$ only has moments of order strictly less than $%
1+\theta $.\newline
\end{corollary}

\textbf{Proof:} If $\theta \in \left( 0,1\right) $, the tail pgf of $%
N_{\infty }$\ is 
\[
\frac{1-\phi _{\infty }\left( z\right) }{1-z}=\left( \alpha +\beta \left(
1-z\right) ^{\theta }\right) ^{-1/\theta }, 
\]
and the proof follows from Tauberian theorem, observing 
\[
\left( \alpha +\beta \left( 1-z\right) ^{\theta }\right) ^{-1/\theta }%
\stackunder{z\downarrow 1}{\sim }\mu _{\infty }\left( 1-\frac{\beta }{\alpha
\theta }\left( 1-z\right) ^{\theta }\right) . 
\]
\newline

$\left( A\right) /\left( ii\right) .$ In the subcritical case $\left(
A\right) /\left( ii\right) $, with $\theta \in \left( -1,1\right] \backslash
\left\{ 0\right\} $, $z_{c}>1$, $a\in \left( 0,1\right) $, $b=\left(
1-a\right) \left( z_{c}-1\right) ^{-\theta }$. Here, with $\phi \left(
1\right) =1$ (a regular case) 
\begin{equation}
\phi \left( z\right) =z_{c}-\left( a\left( z_{c}-z\right) ^{-\theta }+\left(
1-a\right) \left( z_{c}-1\right) ^{-\theta }\right) ^{-1/\theta }\text{ and }%
\mu =\phi ^{\prime }\left( 1\right) =a<1.  \label{phi}
\end{equation}
Let $h\left( z\right) =z_{c}-z=h^{-1}\left( z\right) ,$ $g\left( z\right)
=z^{-\theta }$ and $f\left( z\right) =g\left( h\left( z\right) \right)
=\left( z_{c}-z\right) ^{-\theta }$. The above equation is also \cite{Hop} 
\[
f\left( \phi \left( z\right) \right) =af\left( z\right) +\left( 1-a\right)
f\left( 1\right) . 
\]
Let us look for an invertible function $A\left( z\right) $ with inverse $%
B\left( x\right) =A^{-1}\left( x\right) $ such that $\phi \left( z\right)
=B\left( \mu A\left( z\right) \right) =B\left( aA\left( z\right) \right) $.
Combining the two equations, we should have 
\begin{eqnarray*}
f\circ B\left( aA\left( z\right) \right) &=&af\left( z\right) +\left(
1-a\right) f\left( 1\right) \text{ or} \\
f\circ B\left( ax\right) &=&af\circ B\left( x\right) +\left( 1-a\right)
f\left( 1\right)
\end{eqnarray*}
leading to an affine solution $f\circ B\left( x\right) =\alpha x+\beta $
with $\beta =f\left( 1\right) $ and $\alpha $ left undetermined so far. We
get 
\begin{eqnarray*}
B\left( x\right) &=&f^{-1}\left( \alpha x+f\left( 1\right) \right)
=z_{c}-\left( \alpha x+f\left( 1\right) \right) ^{-1/\theta } \\
A\left( z\right) &=&B^{-1}\left( z\right) =\frac{1}{\alpha }\left( \left(
z_{c}-z\right) ^{-\theta }-\left( z_{c}-1\right) ^{-\theta }\right) .
\end{eqnarray*}
We thus have $\phi _{\infty }\left( z\right) =1-A\left( z\right) =1-\frac{1}{%
\alpha }\left( \left( z_{c}-z\right) ^{-\theta }-\left( z_{c}-1\right)
^{-\theta }\right) $ with $\phi _{\infty }\left( 1\right) =1$. Imposing $%
\phi _{\infty }\left( 0\right) =0$ yields $\alpha =z_{c}^{-\theta }-\left(
z_{c}-1\right) ^{-\theta }$ and so

\begin{proposition}
\begin{equation}
\phi _{\infty }\left( z\right) =1-\left( \frac{\left( z_{c}-z\right)
^{-\theta }-\left( z_{c}-1\right) ^{-\theta }}{z_{c}^{-\theta }-\left(
z_{c}-1\right) ^{-\theta }}\right) =\frac{1-\left( 1-z/z_{c}\right)
^{-\theta }}{1-\left( 1-1/z_{c}\right) ^{-\theta }}  \label{Yag2}
\end{equation}
is the searched pgf of the unique Yaglom limit $N_{\infty }$ in this case
study. It has finite mean $\phi _{\infty }^{\prime }\left( 1\right) $ (and
moments) and $\mathbf{P}\left( N_{\infty }=k\right) $ is asymptotically
equivalent to\ $k^{\theta -1}z_{c}^{-k}$ with both power-law and
geometrically decaying factors.\newline
\end{proposition}

The case $\theta =0$ is finally obtained by continuity.

\begin{corollary}
If $\theta =0$, we get a logarithmic pgf for $N_{\infty }$ as a result of 
\begin{equation}
\phi _{\infty }\left( z\right) =\frac{1-\left( 1-z/z_{c}\right) ^{-\theta }}{%
1-\left( 1-1/z_{c}\right) ^{-\theta }}\stackunder{\left| \theta \right|
\rightarrow 0}{\rightarrow }\frac{-\log \left( 1-z/z_{c}\right) }{-\log
\left( 1-1/z_{c}\right) },  \label{Yag3}
\end{equation}
with mean $\phi _{\infty }^{\prime }\left( 1\right) =-\frac{1}{\left(
z_{c}-1\right) \log \left( 1-1/z_{c}\right) }>1$.\emph{\ }\newline
\end{corollary}

$\left( A\right) /\left( iii\right) .$ In the subcritical case $\left(
A\right) /\left( iii\right) $ (pure death case with $\phi \left( z\right)
=\pi \left( 0\right) +\pi \left( 1\right) z$ and $\mu =\pi \left( 1\right)
<1 $), $N_{n}\mid N_{n}>0\stackrel{d}{\rightarrow }N_{\infty }$ where simply 
$N_{\infty }=1$ whose pgf $\phi _{\infty }\left( z\right) :=\mathbf{E}\left(
z^{N_{\infty }}\right) =z$ clearly obeys the Schr\"{o}der functional
equation 
\[
\overline{\phi }_{\infty }\left( \phi \left( z\right) \right) =\mu \overline{%
\phi }_{\infty }\left( z\right) \text{, }\overline{\phi }_{\infty }\left(
z\right) =1-z. 
\]
Obviously, $\phi \left( z\right) =\overline{\phi }_{\infty }^{-1}\left( \mu 
\overline{\phi }_{\infty }\left( z\right) \right) $ and thus $\phi ^{\circ
n}\left( z\right) =\overline{\phi }_{\infty }^{-1}\left( \mu ^{n}\overline{%
\phi }_{\infty }\left( z\right) \right) =1-\mu ^{n}\left( 1-z\right) $ as
required.\newline

$\bullet $ \textbf{Critical case} \textbf{with }$\mu =1$:

This concerns the case $\left( B\right) $ when $\theta \in \left( 0,1\right] 
$, $z_{c}=1$, $a=1$, $b>0$. We have 
\begin{eqnarray*}
\phi \left( z\right) &=&1-\left( \left( 1-z\right) ^{-\theta }+b\right)
^{-1/\theta } \\
\phi ^{\circ n}\left( z\right) &=&1-\left( \left( 1-z\right) ^{-\theta
}+nb\right) ^{-1/\theta }.
\end{eqnarray*}
This is a regular case with $\phi \left( 1\right) =1$.

\begin{proposition}
The process goes extinct with probability $1$ but it takes a long time to do
so. Indeed, 
\begin{eqnarray*}
\mathbf{P}\left( \tau _{1,0}>n\right) =1-\phi ^{\circ n}\left( 0\right)
=\left( 1+nb\right) ^{-1/\theta }\sim \left( nb\right) ^{-1/\theta }, \\
\mathbf{P}\left( \tau _{i,0}>n\right) =1-\phi ^{\circ n}\left( 0\right)
^{i}\sim i\left( nb\right) ^{-1/\theta },\text{ for large }n\text{,}
\end{eqnarray*}
with persistent heavy tails, non-geometric.\emph{\ }\newline
\end{proposition}

The pgf of $N_{n}\left( 1\right) $ conditioned on $N_{n}\left( 1\right) >0$
is 
\[
\frac{\phi ^{\circ n}\left( z\right) -\phi ^{\circ n}\left( 0\right) }{%
1-\phi ^{\circ n}\left( 0\right) },
\]
therefore 
\begin{eqnarray*}
\mathbf{E}\left( N_{n}\left( 1\right) \mid N_{n}\left( 1\right) >0\right) 
&=&\left( 1+nb\right) ^{1/\theta }\sim b^{1/\theta }n^{1/\theta }, \\
\mathbf{E}\left( N_{n}\left( i\right) \mid N_{n}\left( i\right) >0\right) 
&\sim &ib^{1/\theta }n^{1/\theta },\text{ for large }n\text{,}
\end{eqnarray*}
with slow algebraic growth of order $n^{1/\theta }$ in $n$. A direct
computation shows that 
\[
\phi ^{\prime \prime }\left( z\right) =\frac{b\left( \theta +1\right) \left(
1-z\right) ^{\theta -1}}{\left( 1+b\left( 1-z\right) ^{\theta }\right)
^{1/\theta +2}}.
\]
Because\textbf{\ }$\phi ^{\prime \prime }\left( 1\right) =2b<\infty $ only
when $\theta =1$, it holds (\cite{Harris}, \cite{AN}) that, if $\theta =1$, $%
\mathbf{E}\left( N_{n}\left( 1\right) \mid N_{n}\left( 1\right) >0\right)
\sim nb$ and 
\[
\mathbf{P}\left( \frac{N_{n}\left( 1\right) }{nb}>x\mid N_{n}\left( 1\right)
>0\right) \stackunder{n\rightarrow \infty }{\rightarrow }e^{-x}\text{, }x>0.
\]
\newline

$\bullet $ \textbf{Regular supercritical case} \textbf{with }$\mu <\infty $.

In the supercritical case $\left( C\right) $ for which $\mu =a^{-1/\theta
}<\infty $ ($z_{c}=1$, $\theta \in \left( 0,1\right] $, $a\in \left(
0,1\right) $), $\mu ^{-n}N_{n}\stackrel{d}{\rightarrow }W$ where $W\geq 0$
is a random variable with value in $\Bbb{R}_{+}=\left[ 0,\infty \right) $
whose Laplace-Stieltjes transform $\phi _{W}\left( \lambda \right) :=\mathbf{%
E}\left( e^{-\lambda W}\right) $, $\lambda \geq 0$, obeys the Poincar\'{e}
functional equation 
\begin{equation}
\phi _{W}\left( \mu \lambda \right) =\phi \left( \phi _{W}\left( \lambda
\right) \right) .  \label{Poin}
\end{equation}
Note $\phi \left( z\right) =\phi _{W}\left( \mu \phi _{W}^{-1}\left(
z\right) \right) $ and thus $\phi ^{\circ n}\left( z\right) =\phi _{W}\left(
\mu ^{n}\phi _{W}^{-1}\left( z\right) \right) $.

\begin{proposition}
With $\alpha =\frac{1-a}{a+b-1}>0$ and $\beta =\frac{b}{a+b-1}>0$ ($\beta
-\alpha =1$), if $z_{c}=1$, the Laplace-Stieltjes transform of the
asymptotic growth rate $W$ of $\mu ^{-n}N_{n}$ is 
\begin{equation}
\phi _{W}\left( \lambda \right) =1-\lambda \alpha ^{1/\theta }\left( \beta
\lambda ^{\theta }+1\right) ^{-1/\theta }=\rho +\overline{\rho }\left(
1-\left( 1+\beta ^{-1}\lambda ^{-\theta }\right) ^{-1/\theta }\right) .
\label{Poinsol}
\end{equation}
The extinction probability is $\phi _{W}\left( \infty \right) =\rho
=1-\left( \frac{\alpha }{\beta }\right) ^{1/\theta }$and $W$ has an atom at $%
r=0$ with mass $\rho $. We have $\phi _{W}\left( 0\right) =1$ and the mean
of $W$ is $\mu _{W}:=-\phi _{W}^{\prime }\left( 0\right) =\alpha ^{1/\theta }
$. \newline
\end{proposition}

For general supercritical BGW processes, the limiting $W$ given $W>0$ is
known to be infinitely divisible in some but not all cases \cite{Big}. We
don't know if $W\mid W>0$ here in (\ref{Poinsol}) is infinitely divisible or
not.

\begin{corollary}
If $\theta =1$, 
\[
\phi _{W}\left( \lambda \right) =1-\lambda \alpha \left( \beta \lambda
+1\right) ^{-1}=\frac{\lambda +1}{\beta \lambda +1}=\frac{1}{\beta }+\left(
1-\frac{1}{\beta }\right) \frac{1}{1+\beta \lambda }
\]
is the Laplace-Stieltjes transform of an exponential random variable with an
atom at $0$ with mass $\rho =1/\beta $ and mean $\overline{\rho }\beta =%
\frac{1-a}{a+b-1}=\alpha $. And $\mathbf{P}\left( W>r\mid W>0\right) \sim
e^{-r/\beta }$ decays exponentially fast.\newline
\end{corollary}

Furthermore, using \cite{Feller}, p. $445$,

\begin{corollary}
If $\theta \in \left( 0,1\right) $, $\phi _{W}\left( \lambda \right) \sim
\rho +\overline{\rho }\left( 1-\beta ^{1/\theta }\lambda \right) $ as $%
\lambda $ is close to $0$, meaning exponential tails again, now with $%
\mathbf{P}\left( W>r\mid W>0\right) \stackunder{r\rightarrow \infty }{\sim }%
e^{-r/\beta ^{1/\theta }}$. As $\lambda $ is close to $\infty $, $\phi
_{W}\left( \lambda \right) \sim \rho +\overline{\rho }\left( \beta \theta
\right) ^{-1}\lambda \Sp -\theta  \\  \endSp $, meaning heavy algebraic left
tails $\mathbf{P}\left( W\leq r\mid W>0\right) \stackunder{r\rightarrow 0}{%
\sim }\left( \beta \theta \right) ^{-1}r\Sp \theta  \\  \endSp /\Gamma
\left( 1+\theta \right) $.
\end{corollary}

\subsection{Limit laws (super-critical with infinite mean cases)}

There are two different regimes, depending on $\mu =\infty $ resulting or
not from finite-time explosion:

$\bullet $ \textbf{Regular case.} If $z_{c}=1$, the infinite mean case $\mu
=\infty $ occurs when $\theta =0$, $a\in \left( 0,1\right) $. In such a
case, $\phi \left( z\right) =1-\left( 1-\rho \right) ^{1-a}\left( 1-z\right)
^{a}$ and $\phi \left( 1\right) =1$ (no finite-time explosion). With $%
E\left( 1\right) $ a standard mean $1$ exponential random variable 
\begin{equation}
a^{n}\log \left( 1+N_{n}\left( 1\right) \right) \stackrel{a.s.}{\rightarrow }%
W=\left\{ 
\begin{array}{c}
0\text{ with probability }\rho \\ 
E\left( 1\right) \text{ with probability }\overline{\rho }
\end{array}
\right. ,\text{ as }n\rightarrow \infty  \label{exp}
\end{equation}
and conditionally given $N_{n}\left( 1\right) $ does not go extinct, $%
N_{n}\left( 1\right) $ grows at double exponential speed.

The pgf of $N_{n}\left( 1\right) $ given explosion indeed is 
\[
\phi _{\infty }^{\circ n}\left( z\right) =1-\left( 1-z\right) ^{a^{n}},
\]
and the above statement follows from the martingale proof of \cite{Henard},
proposition $3.8$, adapted to the discrete time context. Similar regular
models with infinite offspring mean were recently studied in \cite{Hui}.%
\newline

\textbf{Remark:} It can be checked that, with $\log _{a}b=\log b/\log a$ and 
$A\left( z\right) =1-\log _{1-\rho }\left( 1-z\right) $, $z<1$, 
\begin{equation}
\phi \left( z\right) =A^{-1}\left( aA\left( z\right) \right) ,\text{ so that 
}\phi ^{\circ n}\left( z\right) =A^{-1}\left( a^{n}A\left( z\right) \right) .
\label{comp}
\end{equation}
This is an alternative way to see that such a branching model is
`integrable'.\newline

$\bullet $ \textbf{Explosive case.} If $\left( i\right) $ $z_{c}>1$ and $%
\theta \in \left( -1,1\right] $, $a\in \left( 0,1\right) $, $b=\left(
1-a\right) \left( z_{c}-\rho \right) ^{-\theta }$ with $\rho \in \left(
0,1\right) $ or $\left( ii\right) $ if $z_{c}=1$ and $\theta \in \left(
-1,0\right) $, $a\in \left( 0,1\right) $ and $b=\left( 1-a\right) \left(
1-\rho \right) ^{-\theta }$, where $\rho \in \left( 0,1\right) $, then $%
N_{n}\left( 1\right) $ can be infinite even in the first iteration step
(finite time explosion). What only matters in this context is the time $\tau
_{1,\infty }$ to explosion and also $\tau _{1}=\tau _{1,0}\wedge \tau
_{1,\infty }$, as well as $\tau _{i}$. We get

\begin{proposition}
$\left( i\right) $ When $\theta \in \left( -1,1\right] $, $z_{c}>1$, $a\in
\left( 0,1\right) $, $b=\left( 1-a\right) \left( z_{c}-\rho \right)
^{-\theta }$ with $\rho \in \left( 0,1\right) $, leading to $\mu =\infty $,
we have for instance 
\[
\mathbf{P}\left( n<\tau _{1}<\infty \right) =\phi ^{\circ n}\left( 1\right)
-\phi ^{\circ n}\left( 0\right) \stackunder{n\rightarrow \infty }{\sim }%
\left( \frac{1-a}{b}\right) ^{1+1/\theta }\left( \left( z_{c}-1\right)
^{\theta }-z_{c}^{\theta }\right) a^{n},
\]
showing that $\tau _{1}$ is tail equivalent to a geometric random variable.
Similarly 
\[
\mathbf{P}\left( n<\tau _{i}<\infty \right) =\phi ^{\circ n}\left( 1\right)
^{i}-\phi ^{\circ n}\left( 0\right) ^{i}\stackunder{n\rightarrow \infty }{%
\sim }i\left( \frac{1-a}{b}\right) ^{1+i/\theta }\left( \left(
z_{c}-1\right) ^{\theta }-z_{c}^{\theta }\right) a^{n}.
\]
$\left( ii\right) $ If $z_{c}=1$ and $\theta \in \left( -1,0\right) $, $a\in
\left( 0,1\right) $ and $b=\left( 1-a\right) \left( 1-\rho \right) ^{-\theta
}$, where $\rho \in \left( 0,1\right) $, we have 
\[
\mathbf{P}\left( n<\tau _{i}<\infty \right) =\phi ^{\circ n}\left( 1\right)
^{i}-\phi ^{\circ n}\left( 0\right) ^{i}\stackunder{n\rightarrow \infty }{%
\sim }-i\frac{\left( 1-\rho \right) ^{-\left( 1+i/\theta \right) }}{\theta }%
a^{n},
\]
still with the tail equivalence to a geometric random variable.
\end{proposition}

\section{Powers of the $\theta $-process transition matrix obtained by
iteration}

So far we dealt with this $\theta $-family of pgfs for the reproduction law.
It remains to compute the probability mass function to which they are
associated. A related question is to compute the stochastic transition
matrix of the $\theta $-branching processes together with its powers in
time. We shall now address these points. We shall start with the cases $%
\theta \in \left( -1,1\right) \backslash \left\{ 0\right\} $ before
addressing the special cases $\theta \in \left\{ -1,0,1\right\} $.

\subsection{The case $\theta \in \left( -1,1\right) \backslash \left\{
0\right\} $}

$\bullet $ We start with \textbf{the reproduction law}. Let $\phi \left(
z\right) =z_{c}-\left( a\left( z_{c}-z\right) ^{-\theta }+b\right)
^{-1/\theta }$ be a $\theta $-pgf with $\phi \left( z\right) \leq 1$. We
first wish to compute the associate probability mass distribution: $\pi
\left( k\right) =\left[ z^{k}\right] \phi \left( z\right) $. Introduce $\phi
_{c}\left( z\right) :=z_{c}^{-1}\phi \left( z_{c}z\right) $, so with $\phi
_{c}\left( z\right) =1-\left( a\left( 1-z\right) ^{-\theta }+bz_{c}^{\theta
}\right) ^{-1/\theta }$ (this operation is meaningful of course only if $%
z_{c}>1$). $\phi _{c}\left( z\right) $ is a new pgf because $\phi _{c}\left(
1\right) =z_{c}^{-1}\phi \left( z_{c}\right) \leq 1$. We have $\pi
_{c}\left( k\right) =\left[ z^{k}\right] \phi _{c}\left( z\right)
=z_{c}^{k-1}\pi \left( k\right) $, so one can work with $\phi _{c}$ as well.
We also have $\phi _{c}\left( z\right) =f\circ g\left( z\right) $ with $%
g\left( z\right) =1-\left( 1-z\right) ^{-\theta }$ and $f\left( z\right)
=1-\left( a+bz_{c}^{\theta }-az\right) ^{-1/\theta }$. This allows to
compute $\pi \left( k\right) $ by Faa di Bruno formula for the composition
of Taylor series. First we have $\pi \left( 0\right) =z_{c}\left( 1-\left(
a+bz_{c}^{\theta }\right) ^{-1/\theta }\right) $. By Faa di Bruno formula (%
\cite{Comtet}, Tome 1, p. $149$), then

\begin{proposition}
\begin{equation}
\pi \left( k\right) =\frac{1}{k!z_{c}^{k-1}}\sum_{l=1}^{k}f_{l}B_{k,l}\left(
g_{\bullet }\right) ,\text{ }k\geq 1,  \label{sprob}
\end{equation}
where $f_{l}$ are the Taylor coefficients of $f\left( z\right) $ and $%
B_{k,l}\left( g_{\bullet }\right) $ the Bell polynomials in the
indeterminate $g_{\bullet }:=\left( g_{1},g_{2},...\right) $, the $g_{k}$s
being the Taylor coefficients of $g\left( z\right) $. The Bell polynomials
are defined by 
\[
B_{k,l}\left( g_{\bullet }\right) =\frac{k!}{l!}\left[ z^{k}\right] g\left(
z\right) ^{l},
\]
with the boundary conditions 
\[
B_{k,0}\left( g_{\bullet }\right) =B_{0,l}\left( g_{\bullet }\right) =0,%
\text{ }k,l\geq 1\text{ and }B_{0,0}\left( g_{\bullet }\right) :=1,\text{
and,}
\]
\[
B_{k,1}\left( g_{\bullet }\right) =g_{k}\text{ and }B_{k,k}\left( g_{\bullet
}\right) =g_{1}^{k}\text{.}
\]
\end{proposition}

This computation of $\pi \left( k\right) $\ is in agreement with Proposition 
$4$\ of Sagitov and Lindo but our representation and its proof, inspired
from Faa di Bruno formulae and making use of Bell polynomials, are
different. We now list some properties concerning the coefficients $f_{l}$\
and $B_{k,l}\left( g_{\bullet }\right) .$ We first recall that \cite{Comtet}%
, 
\[
B_{k,l}\left( x_{\bullet }\right) =k!\stackrel{*}{\sum }\prod_{j\geq 1}\frac{%
1}{c_{j}!}\left( \frac{x_{j}}{j!}\right) ^{c_{j}},
\]
the latter star summations running over the integers $c_{j}$ obeying $%
\sum_{j\geq 1}c_{j}=l$ and $\sum_{j\geq 1}jc_{j}=k\geq l$.

We note now that, with $\left[ a\right] _{l}=a\left( a+1\right) ...\left(
a+l-1\right) $ the ascending factorial with $\left[ a\right] _{0}:=1$, $%
C=\left( a+bz_{c}^{\theta }\right) ^{-1/\theta }$ and $D=\left(
a+bz_{c}^{\theta }\right) /a=C^{-\theta }/a$, $f\left( z\right) =1-C\left(
1-z/D\right) ^{-1/\theta }$ with Taylor coefficients 
\begin{equation}
f_{0}=1-C\text{ and }f_{l}=-\frac{C}{\theta }\left[ 1+1/\theta \right]
_{l-1}D^{-l}=-CD^{-l}\left[ 1/\theta \right] _{l}\text{, }l\geq 1.
\label{fdot}
\end{equation}
For the case $g\left( z\right) =1-\left( 1-z\right) ^{-\theta }$ , it holds $%
g_{\bullet }=-\theta \left[ 1+\theta \right] _{\bullet -1}=-\left[ \theta
\right] _{\bullet }$. Because $g_{1}=-\theta $ and $g_{m+1}=g_{m}\left(
m+\theta \right) $, $m\geq 1$, it follows that the Bell coefficients $%
B_{k,l}\left( g_{\bullet }\right) $ for this function $g$ obey a simple $3-$%
term recursion 
\begin{equation}
B_{k+1,l}\left( g_{\bullet }\right) =-\theta B_{k,l-1}\left( g_{\bullet
}\right) +\left( k+l\theta \right) B_{k,l}\left( g_{\bullet }\right) \text{, 
}k,l\geq 1\text{.}  \label{gdot}
\end{equation}
For instance $B_{1,1}\left( g_{\bullet }\right) =-\theta $ leading to $\pi
\left( 1\right) =f_{1}B_{1,1}\left( g_{\bullet }\right) =CD^{-1}$, $%
B_{2,1}\left( g_{\bullet }\right) =\left( 1+\theta \right) B_{1,1}\left(
g_{\bullet }\right) =-\theta \left( 1+\theta \right) $, leading to $\pi
\left( 2\right) =\frac{1}{2z_{c}}\left( f_{1}B_{2,1}\left( g_{\bullet
}\right) +f_{2}B_{2,2}\left( g_{\bullet }\right) \right) =\frac{1}{2z_{c}}%
\left( 1+\theta \right) CD^{-2}\left( D-1\right) $,$...$The formulae (\ref
{sprob}), (\ref{fdot}) and (\ref{gdot}) completely characterize the $\pi
\left( k\right) $s. The $B_{k,l}\left( g_{\bullet }\right) $ constitute
generalized Stirling numbers studied in \cite{Char}.\newline

\textbf{Remark: }If $\theta =-1/L$ where $L>1$ is an integer, $f\left(
z\right) =1-C\left( 1-z/D\right) ^{L}$ is a polynomial of degree $L$ in $z$
so $f_{l}=0$ if $l>L$ which largely simplifies (\ref{sprob}). Furthermore,
in this case, $g_{k}=L^{-k}\prod_{l=1}^{k-1}\left( lL-1\right) $. If $L=2$, $%
g_{k}=2^{-2\left( k-1\right) }\left( 2k-3\right) !/\left( k-2\right) !$.%
\newline

$\bullet $ \textbf{The transition matrix and its powers}. We now first wish
to compute $P_{a,b}\left( i,j\right) =\left[ z^{j}\right] \phi \left(
z\right) ^{i}=z_{c}^{i-j}\left[ z^{j}\right] \phi _{c}\left( z\right) ^{i}$,
the transition matrix of the $\theta $-branching process, where its
dependence on the parameters $\left( a,b\right) $ has been emphasized. We
have $\phi _{c}\left( z\right) ^{i}=f_{i}\circ g\left( z\right) $, still
with $g\left( z\right) =1-\left( 1-z\right) ^{-\theta }$ and now with $%
f_{i}\left( z\right) :=\left[ 1-\left( a+bz_{c}^{\theta }-az\right)
^{-1/\theta }\right] ^{i}$. So with $f_{i,k}$, $k\geq 1$, the Taylor
coefficients of $f_{i}\left( z\right) $, we similarly get

\begin{proposition}
\begin{equation}
P_{a,b}\left( i,j\right) =\frac{z_{c}^{i-j}}{j!}\sum_{k=1}^{j}f_{i,k}B_{j,k}%
\left( g_{\bullet }\right) .  \label{trans}
\end{equation}
\end{proposition}

We note that $f_{i}\left( z\right) =h_{i}\left( \mathrm{f}\left( z\right)
\right) $ where $h_{i}\left( z\right) =\left( 1-C+Cz\right) ^{i}$ and $%
\mathrm{f}\left( z\right) :=1-\left( 1-z/D\right) ^{-1/\theta }$ so that
with $h_{i,l}=\frac{i!}{\left( i-l\right) !}\left( 1-C\right) ^{i-l}C^{l}$ ($%
=0$ if $l>i$) and with $\mathrm{f}_{\bullet }$ given from (\ref{fdot}) as $%
\mathrm{f}_{l}=-\left[ 1/\theta \right] _{l}D^{-l}$, $l\geq 1$, by Faa di
Bruno formula again 
\begin{equation}
f_{i,0}=\left( 1-C\right) ^{i}\text{ and }f_{i,k}=\sum_{l=1}^{k\wedge
i}h_{i,l}B_{k,l}\left( \mathrm{f}_{\bullet }\right) \text{.}  \label{fs}
\end{equation}
Note $\pi \left( j\right) =P_{a,b}\left( 1,j\right) $ as required.\newline
To obtain now $P_{a,b}^{n}\left( i,j\right) $, the $\left( i,j\right) $%
-entry of the $n$-th power of $P_{a,b},$ we just need to substitute $\left(
a_{n}=a^{n},b_{n}=b\left( 1+a+...+a^{n-1}\right) \right) $ to $\left(
a,b\right) $, so it simply holds 
\begin{equation}
P_{a,b}^{n}\left( i,j\right) =P_{a_{n},b_{n}}\left( i,j\right) ,
\label{powers}
\end{equation}
taking advantage of the invariance under iteration of the $\theta $-family
when $\theta \in \left( -1,1\right) \backslash \left\{ 0\right\} $. We note
that the dependence on $n$ in $P_{a_{n},b_{n}}\left( i,j\right) $ is only in
the coefficients $f_{i,k}$ in (\ref{trans}), through $C$ and $D$. To
emphasize this point, we shall also write

\begin{corollary}
\begin{equation}
P_{a,b}^{n}\left( i,j\right) =P_{a_{n},b_{n}}\left( i,j\right) =\frac{%
z_{c}^{i-j}}{j!}\sum_{k=1}^{j}f_{i,k}^{\left( n\right) }B_{j,k}\left(
g_{\bullet }\right) ,  \label{transn}
\end{equation}
where $f_{i,k}^{\left( n\right) }$ is obtained from $f_{i,k}$ in (\ref{fs})
while substituting $\left( a_{n}=a^{n},b_{n}=b\left( 1+a+...+a^{n-1}\right)
\right) $ to $\left( a,b\right) $ in the expressions of $C=\left(
a+bz_{c}^{\theta }\right) ^{-1/\theta }$ and $D=\left( a+bz_{c}^{\theta
}\right) /a$.\newline
\end{corollary}

It remains to discuss the special integral cases for $\theta $.

\subsection{The case $\theta =0$}

We recall that $\phi \left( z\right) =z_{c}-\lambda \left( z_{c}-z\right)
^{a}$, where $\lambda =\left( z_{c}-\rho \right) ^{1-a}$ and $\rho $ obeys $%
\phi \left( \rho \right) =\rho $. With $\phi _{c}\left( z\right) =1-\lambda
z_{c}^{a-1}\left( 1-z\right) ^{a}$ and $\pi _{c}\left( k\right) =\left[
z^{k}\right] \phi _{c}\left( z\right) $, $\pi \left( k\right) =\pi
_{c}\left( k\right) /z_{c}^{k-1}$ with $\pi _{c}\left( k\right) =-\lambda
z_{c}^{a-1}\left[ -a\right] _{k}/k!$. Next, with $\lambda _{c}:=\lambda
z_{c}^{a-1}$%
\[
\phi _{c}\left( z\right) ^{i}=\left( 1-\lambda _{c}\left( 1-z\right)
^{a}\right) ^{i}=\left( 1-\lambda _{c}+\lambda _{c}\left( 1-\left(
1-z\right) ^{a}\right) \right) ^{i}=h_{i}\circ g\left( z\right) ,
\]
with $g\left( z\right) =1-\left( 1-z\right) ^{a}$ and $h_{i}\left( z\right)
=\left( 1-\lambda _{c}+\lambda _{c}z\right) ^{i}$. With $g_{\bullet
}=-\left[ -a\right] _{\bullet }$ and $h_{i,k}=\frac{i!}{\left( i-k\right) !}%
\left( 1-\lambda _{c}\right) ^{i-k}\lambda _{c}^{k}$, we thus get similarly 
\begin{equation}
P_{a,\lambda }\left( i,j\right) =\frac{z_{c}^{i-j}}{j!}%
\sum_{k=1}^{j}h_{i,k}B_{j,k}\left( g_{\bullet }\right) \text{ and }%
P_{a,\lambda }^{n}\left( i,j\right) =P_{a_{n},\lambda _{n}}\left( i,j\right)
,  \label{powers2}
\end{equation}
where $a_{n}=a^{n}$ ($a\in \left( 0,1\right) $) and $\lambda _{n}=\lambda
^{\left( 1-a^{n}\right) /\left( 1-a\right) }$. The $B_{j,k}\left( g_{\bullet
}\right) $ also obey a three terms recursion of the type (\ref{gdot}) with $%
-a$ substituted to $\theta $. Note $\pi \left( j\right) =P_{a,\lambda
}\left( 1,j\right) $ as required.

\subsection{The case $\theta =1$}

With $\phi \left( z\right) =z_{c}-\left( a\left( z_{c}-z\right)
^{-1}+b\right) ^{-1}$ we wish to compute $\pi \left( k\right) =\left[
z^{k}\right] \phi \left( z\right) $ with $\pi \left( 0\right) =z_{c}\left(
a+b-1\right) /\left( a+b\right) $ in the first place. Introduce $\phi
_{c}\left( z\right) =z_{c}^{-1}\phi \left( z_{c}z\right) $, so with $\phi
_{c}\left( z\right) =1-\left( a\left( 1-z\right) ^{-1}+bz_{c}\right) ^{-1}$.
We have $\phi _{c}\left( z\right) =f\circ g\left( z\right) $ with $g\left(
z\right) :=\left( 1-z\right) ^{-1}-1$ and $f\left( z\right) =1-\left(
a+bz_{c}+az\right) ^{-1}=1-C\left( 1+z/D\right) ^{-1}$ where $C=\left(
a+bz_{c}\right) ^{-1}$ and $D=\left( a+bz_{c}\right) /a$. Let $f_{l}$ be the
Taylor coefficients of $f\left( z\right) $ and $g_{k}$ the Taylor
coefficients of $g\left( z\right) $. By Faa di Bruno formula 
\[
\pi \left( k\right) =\frac{1}{k!z_{c}^{k-1}}\sum_{l=1}^{k}f_{l}B_{k,l}\left(
g_{\bullet }\right) ,\text{ }k\geq 1, 
\]
with $g_{k}=k!$ and $f_{0}=1-C$ and $f_{l}=\left( -1\right) ^{l-1}CD^{-l}l!$%
, $l\geq 1$. We have $B_{k,l}\left( \bullet !\right) =\binom{k-1}{l-1}\frac{%
k!}{l!}$, so 
\begin{equation}
\pi \left( k\right) =\frac{C}{z_{c}^{k-1}}\sum_{l=1}^{k}\binom{k-1}{l-1}%
\left( -1\right) ^{l-1}D^{-l}=CD^{-1}\left( \frac{1-D^{-1}}{z_{c}}\right)
^{k-1}\text{, }k\geq 1.  \label{sprob2}
\end{equation}
Next, 
\[
P_{a,b}\left( i,j\right) =z_{c}^{i-j}\left[ z^{j}\right] \phi _{c}\left(
z\right) ^{i}. 
\]
We have $\phi _{c}\left( z\right) ^{i}=f_{i}\circ g\left( z\right) $ still
with $g\left( z\right) =\left( 1-z\right) ^{-1}-1$ and now with $f_{i}\left(
z\right) =\left[ 1-C\left( 1+z/D\right) ^{-1}\right] ^{i}$. So with $f_{i,k}$%
, $k\geq 1$, the Taylor coefficients of $f_{i}\left( z\right) $ and with $%
B_{j,k}\left( \bullet !\right) =\binom{j-1}{k-1}\frac{j!}{k!}$, we get
similarly 
\begin{equation}
P_{a,b}\left( i,j\right) =\frac{z_{c}^{i-j}}{j!}\sum_{k=1}^{j}f_{i,k}B_{j,k}%
\left( g_{\bullet }\right) .  \label{sto}
\end{equation}
It remains to compute the $f_{i,k}$s. We note that $f_{i}\left( z\right)
=h_{i}\left( f\left( z\right) \right) $ where $h_{i}\left( z\right) =\left(
1-C-Cz\right) ^{i}$ and $f\left( z\right) =\left( 1+z/D\right) ^{-1}-1$ so
that with $h_{i,l}=\frac{i!}{\left( i-l\right) !}\left( 1-C\right)
^{i-l}\left( -C\right) ^{l}$ and with $f_{\bullet }$ given by $f_{l}=\left(
-D\right) ^{-l}l!$, $l\geq 1$, by Faa di Bruno formula again 
\[
f_{i,0}=\left( 1-C\right) ^{i}\text{ and }f_{i,k}=\sum_{l=1}^{k\wedge
i}h_{i,l}B_{k,l}\left( f_{\bullet }\right) \text{.} 
\]
Now, $B_{k,l}\left( f_{\bullet }\right) =\left( -D\right) ^{-k}B_{k,l}\left(
\bullet !\right) =\left( -D\right) ^{-k}\binom{k-1}{l-1}\frac{k!}{l!}$ and 
\begin{equation}
f_{i,k}=k!\left( 1-C\right) ^{i}D^{-k}\sum_{l=1}^{k\wedge i}\left( -1\right)
^{k-l}\binom{i}{l}\binom{k-1}{l-1}\left( \frac{C}{1-C}\right) ^{l}\text{.}
\label{fik}
\end{equation}
Exchanging the summation over $k$ and $l$ in (\ref{sto}) and applying the
binomial identity (keeping in mind $D^{-1}=aC$) 
\begin{equation}
P_{a,b}\left( i,j\right) =z_{c}^{i-j}\left( 1-C\right) ^{i}\left(
1-D^{-1}\right) ^{j}\sum_{l=1}^{i\wedge j}\binom{i}{l}\binom{j-1}{l-1}\left( 
\frac{C}{1-C}\frac{D^{-1}}{1-D^{-1}}\right) ^{l}.  \label{stob}
\end{equation}
\newline

To obtain now $P_{a,b}^{n}\left( i,j\right) $, the $\left( i,j\right) $%
-entry of the $n$-th power of $P_{a,b},$ we just need to substitute $\left(
a_{n}=a^{n},b_{n}=b\left( 1+a+...+a^{n-1}\right) \right) $ to $\left(
a,b\right) $ in $\left( C,D\right) $, so it simply holds 
\[
P_{a,b}^{n}\left( i,j\right) =P_{a_{n},b_{n}}\left( i,j\right) , 
\]
where $P_{a,b}\left( i,j\right) $ is given by (\ref{stob}). The resulting
expression generalizes Proposition $2.2$ of \cite{Kleb}.

\subsection{The case $\theta =-1$ (Greenwood model)}

Here $\phi \left( z\right) =az+z_{c}\left( 1-a\right) -b$. We get $\pi
\left( 1\right) =a$, $\pi \left( 0\right) =z_{c}\left( 1-a\right) -b\leq
1-\pi \left( 1\right) $. We have 
\[
P\left( i,j\right) =\left[ z^{j}\right] \phi \left( z\right) ^{i}=\binom{i}{j%
}\pi \left( 0\right) ^{i-j}\pi \left( 1\right) ^{j}, 
\]
and 
\[
P^{n}\left( i,j\right) =\left[ z^{j}\right] \phi ^{\circ n}\left( z\right)
^{i}=\binom{i}{j}\pi _{n}\left( 0\right) ^{i-j}\pi _{n}\left( 1\right) ^{j}, 
\]
where $\pi _{n}\left( 1\right) =\pi \left( 1\right) ^{n}$ and $\pi
_{n}\left( 0\right) =\pi \left( 0\right) \frac{1-\pi \left( 1\right) ^{n}}{%
1-\pi \left( 1\right) }$. Both $P$ and $P^{n}$ have binomial entries with $%
P^{n}\left( i,i\right) =\pi \left( 1\right) ^{ni}$. If $\pi \left( 0\right)
+\pi \left( 1\right) =1$ (the regular case), $\pi _{n}\left( 0\right) +\pi
_{n}\left( 1\right) =1$ and $P^{n}$ is stochastic. If $\pi \left( 0\right)
+\pi \left( 1\right) <1$ (the explosive case), $\pi _{n}\left( 0\right) +\pi
_{n}\left( 1\right) <1$ and $P^{n}$ is sub-stochastic. To make it
stochastic, we can add state $\left\{ \infty \right\} $ to the state-space
and assume that it is absorbing. We can thus complete $P$ to make it
stochastic while considering $P\left( i,\infty \right) =1-\sum_{j=0}^{i}%
\binom{i}{j}\pi \left( 0\right) ^{i-j}\pi \left( 1\right) ^{j}=1-\left( \pi
\left( 0\right) +\pi \left( 1\right) \right) ^{i}$ and $P\left( \infty
,\infty \right) =1$. If $\phi \left( 1\right) =1$, such regular pure death
process was recently considered by \cite{Moh}, revisiting the Greenwood
model of infectiousness, \cite{Green}.

\subsection{Resolvent of the $\theta $-linear fractional processes}

With $\delta _{i,j}$\ the Kronecker delta, for $i,j\geq 1$, we also obtain
the resolvent of $N_{n}\left( i\right) $ as 
\begin{equation}
g_{i,j}\left( z\right) :=\delta _{i,j}+\sum_{n\geq 1}z^{n}P^{n}\left(
i,j\right) .  \label{resol}
\end{equation}
In particular, 
\[
g_{i,i}\left( z\right) =1+\sum_{n\geq 1}z^{n}P^{n}\left( i,i\right) . 
\]
Note $g_{i,j}\left( 1\right) =\delta _{i,j}+\mathbf{E}\left( \sum_{n\geq 1}%
\mathbf{1}_{\left\{ N_{n}\left( i\right) =j\right\} }\right) $, the expected
value of the time spent on state $j$ starting from $i$, is the Green kernel.

\begin{proposition}
Using (\ref{transn}), with $F_{i,k}\left( z\right) :=\sum_{n\geq
1}z^{n}f_{i,k}^{\left( n\right) }$, we get the following tricky expression
for the resolvent 
\begin{equation}
g_{i,j}\left( z\right) :=\delta _{i,j}+\sum_{n\geq
1}z^{n}P_{a_{n},b_{n}}\left( i,j\right) =\delta _{i,j}+\frac{z_{c}^{i-j}}{j!}%
\sum_{k=1}^{j}F_{i,k}\left( z\right) B_{j,k}\left( g_{\bullet }\right) .
\label{resol2}
\end{equation}
\end{proposition}

These quantities are fundamental to compute pgfs of important quantities
such as passage times. It holds for example that $\mathbf{E}\left( z^{\tau
_{i,j}}\right) =g_{i,j}\left( z\right) /g_{j,j}\left( z\right) $ where 
\begin{equation}
\tau _{i,j}=\inf \left( n\geq 1:N_{n}\left( i\right) =j\right)  \label{hit}
\end{equation}
is the first passage time to state $j\neq i$ of $N_{n}$ given $N_{0}=i$, 
\cite{Norris}, \cite{Woess}. In particular $\mathbf{P}\left( \tau
_{i,j}<\infty \right) =g_{i,j}\left( 1\right) /g_{j,j}\left( 1\right) $ are
the hitting probabilities of state $j$ starting from $i$. Furthermore, with 
\begin{equation}
\tau _{i,i}^{*}=\inf \left( n\geq 1:N_{n}\left( i\right) =i\right) ,
\label{return}
\end{equation}
the first return time to state $i$ of $N_{n}\left( i\right) $, it holds by
renewal arguments that $\mathbf{E}\left( z^{\tau _{i,i}^{*}}\right)
=1-1/g_{i,i}\left( z\right) $, \cite{Norris}. In particular $\mathbf{P}%
\left( \tau _{i,i}^{*}<\infty \right) =1-1/g_{i,i}\left( 1\right) $.
Therefore for example, the mean return time to state $i$ given $\tau
_{i,i}^{*}<\infty $ is 
\begin{equation}
\mathbf{E}\left( \tau _{i,i}^{*}\mid \tau _{i,i}^{*}<\infty \right) =\frac{%
g_{i,i}^{\prime }\left( 1\right) }{g_{i,i}\left( 1\right) \left(
g_{i,i}\left( 1\right) -1\right) },  \label{return2}
\end{equation}
whenever this quantity exists.\newline

Let us briefly sketch what this says for the simplest Greenwood model
example when $\theta =-1$: firstly $P^{n}\left( i,i\right) =\pi _{n}\left(
1\right) ^{j}$ leading to $g_{i,i}\left( z\right) =1+\sum_{n\geq 1}z^{n}\pi
\left( 1\right) ^{ni}=1/\left( 1-z\pi \left( 1\right) ^{i}\right) $.
Therefore $\mathbf{E}\left( z^{\tau _{i,i}^{*}}\right) =z\pi \left( 1\right)
^{i}$, translating the fact that $\tau _{i,i}^{*}=1$ with probability $\pi
\left( 1\right) ^{i}$, $=\infty $ with probability $1-\pi \left( 1\right)
^{i}$ (the no return to $i$ event if in the first step one of the $i$
founders moved to one of the absorbing states, $0$ or $\infty $). In
addition, in the regular case $\pi \left( 0\right) =1-\pi \left( 1\right) $, 
\[
\mathbf{P}\left( \tau _{i,j}<\infty \right) =g_{i,j}\left( 1\right)
/g_{j,j}\left( 1\right) =\left( 1-\pi \left( 1\right) ^{j}\right) \left(
1+\sum_{n\geq 1}\binom{i}{j}\left( 1-\pi \left( 1\right) ^{n}\right)
^{i-j}\pi \left( 1\right) ^{nj}\right) , 
\]
which, upon developing $\left( 1-\pi \left( 1\right) ^{n}\right) ^{i-j}$ and
summing over $n$ is Proposition $1.1$ and Theorem $1.2$ of \cite{Moh}.

\section{One illustrative example}

As an illustrative application of the previous results, let us look for the
value of $n$ for which a supercritical process as in $\left( C\right) $ will
nearly never (with large probability $c$) go extinct as soon as $N_{n}\left(
i\right) >0$. It is given by (\ref{neverext}) 
\[
c=\frac{1-\rho ^{i}}{1-\phi ^{\circ n}\left( 0\right) ^{i}}=:1-\epsilon . 
\]
When $\epsilon $ is small, it leads to 
\begin{equation}
\rho -\phi ^{\circ n}\left( 0\right) \approx \epsilon \cdot \frac{1-\rho ^{i}%
}{i\rho ^{i-1}}.  \label{asymp}
\end{equation}
The condition for this Taylor expansion to be valid is given\footnote{%
It is assumed here that $i>1$. If $i=1$, the condition on $\rho-\phi^{\circ
n} \left( 0\right)$ is no longer valid, but the one on $\epsilon$ still is.}
by $\rho -\phi ^{\circ n}\left( 0\right) \ll \rho /(i-1)$ or alternatively 
\[
\epsilon \ll \frac{\rho ^{i}}{1-\rho ^{i}}. 
\]
The relation (\ref{asymp}) shows that for small $\epsilon $, having $i$
founders amounts simply to multiply $\epsilon $ by a factor that only
depends on the value of the fixed point $\rho $ and the number $i$. Let us
now use the explicit form of the supercritical $\theta $-linear fractional
pgfs. In that case, it holds that 
\[
a^{n}=\frac{\left( z_{c}-\phi ^{\circ n}\left( z\right) \right) ^{-\theta
}-\left( z_{c}-\rho \right) ^{-\theta }}{\left( z_{c}-z\right) ^{-\theta
}-\left( z_{c}-\rho \right) ^{-\theta }}. 
\]
A Taylor expansion of $\left( z_{c}-\phi ^{\circ n}\left( z\right) \right)
^{-\theta }$ for small $\rho -\phi ^{\circ n}\left( 0\right) $ yields 
\[
-\theta \frac{\rho -\phi ^{\circ n}\left( z\right) }{\rho -z_{c}}\approx
a^{n}\left( 1-\left( \frac{z_{c}-z}{z_{c}-\rho }\right) ^{-\theta }\right) , 
\]
the Taylor expansion validity condition being $\left| \rho -\phi ^{\circ
n}\left( z\right) \right| \ll \left| (z_{c}-\rho )/\theta \right| $.
Combined with the previous result, we get 
\[
a^{n}\approx \epsilon \cdot \frac{1-\rho ^{i}}{i\rho ^{i}}\cdot (-\theta )%
\frac{1-\alpha }{1-\alpha ^{-\theta }}, 
\]
with $\alpha :=z_{c}/(z_{c}-\rho )>1$; the validity conditions are 
\begin{equation}
\epsilon \ll \frac{\rho ^{i}}{1-\rho ^{i}}\text{ and }\epsilon \ll \frac{%
i\rho ^{i}}{\left| \theta \right| \left( \alpha -1\right) }.  \label{cond}
\end{equation}
This shows that the searched value of $n$ for which a supercritical process
as in $\left( C\right) $ will nearly never go extinct is approximately the
sum of three terms:

\begin{itemize}
\item  one, related to the required accuracy, that is the logarithm of $%
\epsilon $ in base $a$. In particular, to have a result $10$ times more
precise, one has to wait $\left| \log _{a}(10)\right| $ more steps,

\item  one, related to the number $i$ of founders, which also depends on the
parameters $a$ and $\rho $,

\item  one, related to the model parameters only, which depends on $a$, $%
\theta $, and $z_{c}$ and $\rho $ through $\alpha $.\newline
\end{itemize}

For instance, taking $a=0.63$ (so that $a^{5}\approx 0.1$), $\rho =0.7$, $%
z_{c}=1$, we get:

\begin{itemize}
\item  when $\theta =+1$ and $i=1$, $9$ generations are needed if the
population is to survive with a probability $1-10^{-2}$. $5$ more
generations will increase this probability to $1-10^{-3}$, and another $5$
to $1-10^{-4}$.

\item  when $\theta =+1$, with an uncertainty $\epsilon =10^{-4}$ and eight
founders, the time to wait decreases to $16$ generations. With thirteen
founders, it decreases further to $13$ generations. Notice that $\rho
^{19}\approx 0.001$, so one has to be careful not to get out of the range of
(\ref{cond}).

\item  with one founder and an uncertainty $10^{-4}$, $19$ generations are
needed for $\theta =1$, $20$ generations for the limit $\theta =0$ and $21$
for $\theta =-1$.\newline
\end{itemize}

In all these special cases, we conclude that if extinction is to occur, it
occurs rapidly or nearly never.\newline

\textbf{Acknowledgments:}

T. Huillet acknowledges partial support from the ``Chaire \textit{%
Mod\'{e}lisation math\'{e}matique et biodiversit\'{e}''.} N. Grosjean and T.
Huillet also acknowledge support from the labex MME-DII Center of Excellence
(\textit{Mod\`{e}les math\'{e}matiques et \'{e}conomiques de la dynamique,
de l'incertitude et des interactions}, ANR-11-LABX-0023-01 project).

\end{document}